\documentclass[12pt,showpacs,amssymb,superscriptaddress]{revtex4}
\usepackage{mathrsfs}
\newtheorem{lem}{Lemma}
\newtheorem{prop}{Proposition}
\newtheorem{theo}{Theorem}

\begin{document}

\title{Quantum Gowdy $T^3$ model: A uniqueness result}

\author{Alejandro Corichi}\email{corichi@matmor.unam.mx}
\affiliation{Instituto de Matem\'aticas, Universidad Nacional
Aut\'onoma de M\'exico, A. Postal 61-3, Morelia, Michoac\'an 58090,
Mexico.} \affiliation{Instituto de Ciencias Nucleares, Universidad
Nacional Aut\'onoma de M\'exico, A. Postal 70-543, M\'exico D.F.
04510, Mexico.}
\author{ Jer\'onimo Cortez}\email{jacq@iem.cfmac.csic.es}
\affiliation{Instituto de Estructura de la Materia, CSIC, Serrano
121, 28006 Madrid, Spain.}
\author{Guillermo A. Mena Marug\'an}\email{mena@iem.cfmac.csic.es}
\affiliation{Instituto de Estructura de la Materia, CSIC, Serrano
121, 28006 Madrid, Spain.}
\author{Jos\'e M. Velhinho}\email{jvelhi@ubi.pt}
\affiliation{Departamento de F\'{\i}sica, Universidade da Beira
Interior, R. Marqu\^es D'\'Avila e Bolama, 6201-001 Covilh\~a,
Portugal.}

\begin{abstract}
Modulo a homogeneous degree of freedom and a global constraint, the
linearly polarised Gowdy $T^3$ cosmologies are equivalent to a free
scalar field propagating in a fixed nonstationary background.
Recently, a new field parameterisation was proposed for the metric
of the Gowdy spacetimes such that the associated scalar field
evolves in a flat background in 1+1 dimensions with the spatial
topology of $S^1$, although subject to a time dependent potential.
Introducing a suitable Fock quantisation for this scalar field, a
quantum theory was constructed for the Gowdy model in which the
dynamics is implemented as a unitary transformation. A question that
was left open is whether one might adopt a different, nonequivalent
Fock representation by selecting a distinct complex structure. The
present work proves that the chosen Fock quantisation is in fact
unique (up to unitary equivalence) if one demands unitary
implementation of  the dynamics and invariance under the group of
$S^1$-translations. These translations are precisely those generated
by the global constraint that remains on the Gowdy model. It is also
shown that the proof of uniqueness in the choice of complex
structure can be applied to more general field dynamics than that
corresponding to the Gowdy cosmologies.

\vskip 3mm \noindent
\end{abstract}

\maketitle
\newpage
\renewcommand{\thefootnote}{\fnsymbol{footnote}}

\section{Introduction}
\label{int}

The quantisation of systems which possess fieldlike degrees of
freedom involves choices that generally lead to inequivalent
theories within the standard Hilbert space approach \cite{wald}.
Opposite to the situation found for systems with a finite
dimensional linear phase space, where the Stone-von Neumann theorem
guarantees that any two strongly continuous, irreducible and unitary
representations of the Weyl relations are unitarily equivalent
\cite{SVN}, in quantum field theory no general uniqueness theorem
can be invoked. Therefore, additional criteria are needed to select
a preferred representation of the canonical commutation relations.
For instance, in background independent quantum gravity
\cite{ash-lew-rep,rov,thie}, the requirement of spatial
diffeomorphism invariance provides a unique representation of the
kinematical holonomy-flux algebra  \cite{lost}. For field theories
in Minkowski spacetime, the criterion of Poincar\'e invariance is
naturally employed to arrive at a unique representation. In
particular, if the field theory corresponds to a Klein-Gordon field,
Poincar\'e invariance selects a complex structure (which is the
mathematical object that encodes the ambiguity in the quantisation)
and thus picks out a preferred representation of the Weyl relations
\cite{poincare}. In fact, even when the Klein-Gordon field
propagates in a more general but still stationary spacetime, a
preferred complex structure can be selected by imposing the energy
criterion introduced in \cite{ash-mag}. In spite of these examples,
it should be emphasised that, in generic curved spacetimes, no
uniqueness criteria exist and field theories generally admit
infinitely many unitarily inequivalent representations.

In the framework of canonical quantum gravity, symmetry reduced
models have been very useful to discuss conceptual and technical
issues in a concrete arena. Though the reduction is drastic for the
so-called minisuperspace models \cite{misner1}, in the sense that
only a finite number of degrees of freedom remain in the system,
midisuperspace models \cite{midi-sm} still retain the field
complexity of general relativity after symmetry reduction,
possessing local degrees of freedom. As a consequence,
midisuperspace models have to face the inherent ambiguity that is
associated with the quantisation of fields. In particular, in order
to deal with their quantisation, one has to address the question of
the unitary equivalence of representations and investigate whether
there exist physical criteria that select a preferred one. In the
present work, we shall analyse the uniqueness of the Fock
quantisation of a particularly interesting midisuperspace model,
namely, the model introduced in \cite{cocome2,cocome} for the
description of the linearly polarised Gowdy $T^3$ cosmologies
\cite{gowdy}.

These cosmologies provide the simplest of all inhomogeneous
(spatially closed) cosmological systems. They are vacuum spacetimes
whose spatial sections have the topology of a three-torus and which
possess two commuting, spacelike and hypersurface orthogonal Killing
vector fields \cite{gowdy}. As a midisuperspace model, this family
of cosmologies has local gravitational degrees of freedom that are
described just by one scalar field. The classical solutions
correspond to spacetimes with a big-bang singularity. Therefore, the
model supplies a nontrivial cosmological scenario where one can
study fundamental questions about canonical quantum gravity and
quantum field theory in curved spacetime. This explains the interest
that has been paid in the literature to its quantisation
\cite{cocome2,cocome,misn,berger,hs,guillermo,pierri,ccq-t3,come,
torre-prd}.

The first preliminary attempts to construct a quantum version of
these cosmologies and obtain physical predictions date back to the
seventies \cite{misn,berger}. The problem was later revisited within
the nonperturbative quantisation framework by using Ashtekar
variables \cite{hs,guillermo}. Nonetheless, it was only recently
that real progress was achieved in defining a complete quantisation
\cite{pierri}. However, it was almost immediately noticed that, in
the Fock quantisation put forward in \cite{pierri}, the dynamics
cannot be implemented as a unitary transformation, neither on the
kinematical \cite{ccq-t3,come} nor on the physical \cite{torre-prd}
Hilbert space. This failure of unitarity precludes the availability
of a Schr\"{o}dinger picture with a conventional notion of
probability preserved by the evolution \cite{torre-prd,jacob}. In
order to reconcile the quantisation of the linearly polarised Gowdy
model with the standard probabilistic interpretation of quantum
physics, an alternative nonperturbative canonical quantisation of
the model was recently proposed in \cite{cocome2,cocome}. In this
new Fock quantisation, the dynamics is indeed unitary. In this
sense, this is the first available example of a consistent
quantisation of an inhomogeneous cosmological system.

In the description of the model proposed in \cite{cocome2}, the
quantisation of the local gravitational degrees of freedom is
obtained by exploiting the equivalence of the classical solutions
with those corresponding to a real scalar field in a fictitious
background. More precisely, once the classical system has been
(almost completely) gauge fixed and a choice of internal time has
been made, the spacetimes are characterised, modulo a remaining
global constraint, by a ``point particle" degree of freedom and by a
real scalar field $\xi$. The point particle degree of freedom has a
trivial evolution (it corresponds to a canonical pair of constants
of motion). Moreover, it plays no role in the discussion of the
uniqueness of the quantisation, since this only affects systems with
an infinite number of degrees of freedom. We shall hence restrict
our analysis from now on to the field sector of the model.

The field $\xi$ is subject to a time dependent potential
$V(\xi)=\xi^2/(4t^2)$ and propagates in a fictitious flat spacetime
in 1+1 dimensions whose spatial slices have circular
topology$^{\footnotemark[5]}$,\footnotetext[5]{One may alternatively
consider an axially symmetric field propagating in a 2+1 dimensional
background with the spatial topology of a two-torus,
${\cal{M}}\approx {\mathbb{R}}^{+}\times T^2$ \cite{cocome2,cocome}.
For simplicity, we here adopt the 1+1 dimensional perspective.}
${\cal{M}}\approx {\mathbb{R}}^{+}\times S^1$. This evolution is
governed by the time dependent Hamiltonian $^{\footnotemark[6]}$
\footnotetext[6]{We employ a system of units with $c=4G/\pi=1$, $c$
and $G$ being the speed of light and Newton's constant,
respectively.}
\begin{equation} \label{ham} H={1\over 2}\oint
\left[P_{\xi}^2+(\xi')^2+ {\xi^2\over 4t^2}\right] d\theta,
\end{equation}
where $\theta\in S^1$ is the spatial coordinate, $t\in
\mathbb{R}^{+}$ is the time coordinate and the prime denotes the
derivative with respect to $\theta$. In addition, $P_{\xi}$ is the
momentum canonically conjugate to $\xi$. So, the symplectic
structure $\Omega$ on the field sector of the canonical phase space
is
\begin{equation}\label{symcanoxi}
\Omega([\xi_{1},P_{\xi_1}],[\xi_{2},P_{\xi_2}])=\oint \left(\xi_{2}
P_{\xi_1}-\xi_{1} P_{\xi_2}\right) d\theta. \end{equation} The
Hamiltonian equations of motion are then
\begin{equation} \label{2} \dot \xi=P_{\xi},\quad \quad
\dot P_{\xi}=\xi''-{\xi\over 4t^2}, \end{equation} where the dot
stands for the derivative with respect to $t$. Thus, in agreement
with our previous comments, the field satisfies the second order
equation
\begin{equation} \label{3} \ddot\xi-\xi''+{\xi\over 4t^2}=0.
\end{equation}

It is worth noticing that, since the Hamiltonian does not depend on
$\theta$, the field equations are invariant under constant
$S^1$-translations,
\begin{equation} \label{4} T_{\alpha}: \theta\mapsto\theta+\alpha
\quad\quad \forall\alpha\in S^1.
\end{equation}
Furthermore, in the present case the translations $T_{\alpha}$
$^{\footnotemark[7]}$ \footnotetext[7]{In the following, we will
obviate the word ``constant'' when referring to these translations,
understanding that the angle $\alpha$ is independent of the
spacetime position.} are in fact gauge symmetries, because the
system is subject to a global constraint which is precisely their
generator \cite{cocome2,cocome}:
\begin{equation} \label{const} C_0=\frac{1}{\sqrt{2\pi}}\oint
P_{\xi} \xi^{\prime} d\theta=0.\end{equation}

The quantisation of the field sector of the model reduces then to a
quantum theory of the scalar field $\xi$ in the above mentioned flat
background. The quantum Gowdy model introduced in
\cite{cocome2,cocome} is defined by using a representation for $\xi$
on a fiducial Fock space, resulting in a unitary implementation of
the dynamics as well as of the gauge group of $S^1$-translations.
This automatically provides a quantisation of the global constraint
$C_0$, since this is the generator of the group of translations (and
we consider exclusively weakly continuous unitary implementations of
this group). The constraint can then be imposed to obtain the
physical Hilbert space. One can show that the dynamics is also
unitarily implemented on this space of quantum physical states
\cite{cocome2,cocome}.

In order to arrive at the quantum theory obtained in
\cite{cocome2,cocome} for the Gowdy model, three important choices
are made that may affect the final outcome \cite{cocome2,cocome}.
The first one is the choice of deparameterisation, owing to the
compact nature of the spatial sections. This choice introduces a
fictitious (internal) time that provides the notion of time
evolution. In spite of the inherent ambiguity in this choice, the
time selected is certainly the most natural candidate, since it
corresponds to the square root of the determinant of the metric
induced on the group orbits that are spanned by the two Killing
vectors, and the timelike character of the gradient of this function
is invariant under coordinate transformations \cite{macca}. The
second one is the choice of a field parameterisation for the spatial
metric, which results in the freedom to perform time dependent
canonical transformations of the field $\xi$ and its momentum after
the deparameterisation of the system \cite{cocome}. We assume that
this field parameterisation has been fixed (at least as far as time
dependent canonical transformations are concerned). The consequences
of adopting other field parameterisations will be analysed
elsewhere.

Once the above choices have been made, the quantisation put forward
in \cite{cocome2,cocome} is of the Fock type, i.e. the GNS state
that defines the representation of the kinematical Weyl algebra is
defined by a Hilbert space structure in phase space (or in the space
of smooth solutions), which in turn is uniquely defined by a complex
structure. This is the third choice that may affect the
quantisation. Although the chosen complex structure is a natural
candidate and endows the quantisation with amenable properties, the
question arises of whether a different selection of complex
structure might lead to a different, (unitarily) nonequivalent
quantisation which could still be physically acceptable. This is the
issue that we shall investigate in the present work. We shall show
that, under reasonable requirements, the quantisation put forward in
\cite{cocome2,cocome} is unique.

In particular, these requirements concern the unitary implementation
of the group of gauge transformations (\ref{4}). Since this group
can be implemented in a natural invariant way, i.e. there are states
of the Weyl algebra that are invariant under translations, we
restrict our discussion exclusively to such states and the
corresponding representations. So, we consider only Fock
representations for which the group of $S^1$-translations belongs to
the unitary group of the one-particle Hilbert space. This amounts to
restricting one's attention to complex structures that are left
invariant under those translations.

Our result is thus that  any Fock quantisation defined by a
translation invariant complex structure that provides a unitary
implementation of the dynamics is unitarily equivalent to that
proposed in \cite{cocome2,cocome}.

In addition, we shall see that our proof of uniqueness of the Fock
quantisation may actually be extended to more general dynamics than
the one corresponding to the real scalar field $\xi$ in the
case of the Gowdy model. For instance, the proof is valid for a free
massless field propagating in the same flat background in 1+1
dimensions.

The rest of the paper is organised as follows. Section 2 summarises
the quantisation of the Gowdy model introduced in \cite{cocome2,
cocome} to attain a unitary implementation of the dynamics
$^{\footnotemark[8]}$ \footnotetext[8]{Another midisuperspace where
unitarity problems have been found are the linearly polarised
cylindrical waves \cite{choma}. Actually the detected problems,
which affect the implementation of radial diffeomorphisms, can be
solved in a way similar to that explained for the Gowdy model in
\cite{cocome2}, though changing the roles of the time and spatial
coordinates (namely, by scaling the fundamental scalar field by a
function of the radial coordinate). It would be interesting to see
whether the uniqueness of the corresponding Fock quantisation
results from the demand of unitarity on time evolution and radial
diffeomorphisms, generalising the present analysis to the context of
parameterised field theory.} and introduces the notation that will
be employed in our discussion. In section 3 we determine the complex
structures that are invariant under $S^1$-translations and show that
they are all related by a specific family of symplectic
transformations. Section 4 contains the proof of the uniqueness of
the invariant complex structure (up to unitary transformations of
the Fock representation) under the requirement that the dynamics
admit a unitary implementation. This proof is not restricted to the
case of the Gowdy model, but applies to a broader class of field
dynamics satisfying certain conditions. In section 5 we show that
such conditions are indeed fulfilled by the field evolution
corresponding to the linearly polarised Gowdy cosmologies. We
present our conclusions and some further comments in section 6.
Finally, in the appendix we give alternative uniqueness criteria,
imposing the stronger requirement of a well defined action of the
Hamiltonian on the vacuum of the Fock representation, instead of the
unitary implementation of the dynamics.

\section{The quantum Gowdy model}
\label{sec2}

In this section we shall briefly review the Fock quantisation of the
linearly polarised Gowdy $T^3$ model that was constructed in
\cite{cocome2,cocome}, emphasising those aspects that will be
important for our analysis.

By exploiting the periodicity in the spatial coordinate $\theta$, we
first expand the canonical fields $\xi$ and $P_{\xi}$ in Fourier
series:
\begin{equation} \label{5} \xi(\theta,t)=\sum_{n=-\infty}^{\infty}
\xi_{n}(t){e^{in\theta}\over\sqrt{2\pi}}, \quad\quad
P_{\xi}(\theta,t)=\sum_{n=-\infty}^{\infty}
P^{n}_{\xi}(t){e^{in\theta}\over\sqrt{2\pi}}.
\end{equation} Note that the (implicitly time dependent) Fourier
coefficients $\xi_n$ and $P_{\xi}^{-n}$ are canonically conjugate
and that $\xi_n^{*}=\xi_{-n}$ and $(P_{\xi}^{n})^{*}=P_{\xi}^{-n}$
because the scalar field $\xi(\theta,t)$ and its momentum are real.
The symbol $*$ denotes complex conjugation. Since neither the
unitary implementation of the dynamics and gauge group (\ref{4}), on
the one hand, nor the unitary equivalence of the different
representations, on the other hand, depend on a finite number of
degrees of freedom, we shall obviate the zero mode in the following
for convenience. For the rest of modes we introduce the set of
complex phase space coordinates
\begin{equation}\label{bcal}
\{B_m=( b_m, b_{-m}^*, b_{-m}, b_{m}^*),\, m\in \mathbb{N}\}
\end{equation}
which are given by \begin{equation} \label{6}
b_m={m\xi_{m}+iP_{\xi}^{m}\over \sqrt{2m}},  \quad\quad
b_{-m}^*={m\xi_{m}-iP_{\xi}^{m}\over \sqrt{2m}},
\end{equation}
whereas $b_{-m}$ and $b_{m}^{*}$ are the complex conjugate of
$b_{-m}^{*}$ and $b_{m}$, respectively. The coordinates
$(b_m,b_m^{*})$ and $(b_{-m},b_{-m}^{*})$ are pairs of
annihilationlike and creationlike variables. Here, $\mathbb{N}$ is
the set of all strictly positive integers. In the following, we
shall treat $B_m$ as a column vector for each $m\in \mathbb{N}$. It
is worth pointing out that in the definition of this vector we have
adopted a slightly different order than that employed for the
similar vector ${\cal B}_m$ in \cite{cocome}. The order chosen here
will simplify our expressions.

The above variables have very simple transformation properties under
the translations $T_{\alpha}$, namely
\begin{eqnarray} \label{8} b_m&\mapsto & e^{im\alpha}b_m, \quad
\quad \quad b_{-m}^{*}\mapsto e^{im\alpha}b_{-m}^{*},
\\ b_{-m}&\mapsto & e^{-im\alpha}b_{-m}, \quad \quad
b_m^{*}\mapsto e^{-im\alpha}b_m^{*}. \label{8b}
\end{eqnarray}

On the other hand, as explained in \cite{cocome2,cocome}, the
evolution from $\{B_m(t_0)\}$ at time $t_0$ to $\{B_m(t)\}$ at time
$t$ is determined by a classical evolution operator $U(t,t_0)$ that
has the block diagonal form:
\begin{eqnarray} \label{9} B_m(t)&=&U_m(t,t_0)
B_m(t_0),\\ \label{10} U_m(t,t_0)&=& W(x_m)W(x^0_m)^{-1},
\end{eqnarray} where $x_m=mt$, $x^0_m=mt_0$ and
\begin{equation}
\label{11} W(x)= \left(
\begin{array}{cc}
{\cal W}(x) & {\bf 0}  \\
{\bf 0} & {\cal W}(x)
\end{array} \right),\quad\quad {\cal W}(x)= \left(
\begin{array}{cc}
c(x) & d(x) \\
d^*(x) & c^*(x)
\end{array} \right),\end{equation}
\begin{eqnarray}
\label{12} c(x)&=&\sqrt{{\pi x\over 8}}
\left[ \left(1+\frac{i}{2x}\right)H_0(x)-iH_1(x)\right], \\
\label{13} d(x)&=&\sqrt{{\pi x\over 8}}\left[
\left(1+\frac{i}{2x}\right)H_0^*(x)-iH_1^*(x)\right].
\end{eqnarray} Here, the symbol ${\bf 0}$ denotes the zero
$2\times 2$ matrix, while $H_0$ and $H_1$ are the zeroth and first
order Hankel functions of the second kind, respectively \cite{abra}.
Note that $|c(x)|^2 - |d(x)|^2=1$, so that the map (\ref{9}) is a
Bogoliubov transformation.

The classical evolution matrices (\ref{10}) take then the expression
\begin{equation} \label{14} U_m(t,t_0)= \left(
\begin{array}{cc} {\cal U}_m(t,t_0) & {\bf 0} \\ {\bf 0} &
{\cal U}_m(t,t_0)
\end{array} \right),\quad  {\cal U}_m(t,t_0)
=\left( \begin{array}{cc}\alpha_m(t,t_0) & \beta_m(t,t_0) \\
\beta^*_m(t,t_0) & \alpha^*_m(t,t_0)
\end{array} \right),
\end{equation}
with
\begin{eqnarray} \label{15}
\alpha_m(t,t_0)&=&c(x_m)c^*(x^0_m)-d(x_m)d^*(x^0_m), \\
\label{16} \beta_m(t,t_0)&=&d(x_m)c(x^0_m)-c(x_m) d(x^0_m).
\end{eqnarray}

Finally, in the coordinates $\{ B_m\}$, the symplectic form can also
be decomposed in blocks,
\begin{eqnarray} \label{symstru}
&&\Omega(\{B_m^{(1)}\},\{B_{\tilde{m}}^{(2)}\})=\sum_m (B_m^{(1)})^T
\Omega_m B_m^{(2)},\\ \label{omegam} &&\Omega_m =\left(
\begin{array}{cc} {\bf 0} & \omega_{m} \\ \omega_{m} & {\bf 0}
\end{array} \right), \quad \quad
\omega_{m}= \left(\begin{array}{cc} 0 & -i \\ i & 0 \end{array}
\right),
\end{eqnarray}
where $(B_m^{(1)})^T$ is the row vector transpose of $B_m^{(1)}$.

It is worth noticing at this stage that expressions (\ref{9}) and
(\ref{14}) are not specific of the considered Gowdy model. They are
in fact generic for systems whose classical evolution operator
commutes with the action of the $S^1$-translations and the
$\theta$-reversal transformation $b_m\leftrightarrow b_{-m}$. Of
course, the functions $\alpha_m(t,t_0)$ and $\beta_m(t,t_0)$ are
model dependent. For instance, $\alpha_m(t,t_0)=e^{-im(t-t_0)}$ and
$\beta_m(t,t_0)=0\ \forall m\in\mathbb{N}$ in the case of the free
massless scalar field.

In order to obtain a Fock quantisation of the system, one must now
introduce a complex Hilbert space structure in phase space. This is
done by choosing a complex structure $J$ which, together with the
symplectic form, defines the real part of the inner product
\cite{poincare,fock}. The imaginary part of this inner product
is determined by the symplectic form itself. The specific complex
structure $J_0$ chosen in \cite{cocome2,cocome} is given in the
$\{B_m\}$ basis by a block diagonal matrix, where each $4\times 4$
block has the form $(J_0)_m={\rm diag}(i,-i,i,-i)$.

Let ${\cal H}_0$ be the one-particle Hilbert space determined by
$J_0$, ${\cal F}({\cal H}_0)$ the corresponding (symmetric) Fock
space and $|0\rangle$ the standard cyclic vector (i.e. the vacuum or
zero-particle state). The variables $\{B_m\}$ are precisely those
promoted to the creation and annihilation operators of the Fock
representation defined by $J_0$. In particular, the vacuum is
characterised by the equations
$\hat{b}_m|0\rangle=\hat{b}_{-m}|0\rangle=0$ $\forall m\in
\mathbb{N}$. From definitions (\ref{6}), the complex structure $J_0$
can then be understood as the natural one corresponding to a free
massless dynamics for the scalar field $\xi$ in our flat background.

Furthermore, since the chosen complex structure $J_0$ is invariant
under the group of translations $T_{\alpha}$, one obtains an
invariant unitary implementation of this gauge group, so that
$\forall \alpha\in S^1$ there exists a unitary operator
$\hat{T}_{\alpha}$ such that
\begin{equation} \label{17} \hat{T}_{\alpha}
\hat{b}_m\hat{T}_{\alpha}^{-1}= \widehat{T_\alpha
b_m}=e^{im\alpha}\hat{b}_m \end{equation} and
\begin{equation} \label{18} \hat{T}_{\alpha}|0\rangle=|0\rangle.
\end{equation}
Most importantly, it was proved in \cite{cocome2,cocome} that the
dynamics is also unitarily implementable in this Fock
representation, namely, there are unitary operators $\hat{
U}(t,t_0)$ which satisfy
\begin{equation} \label{19} \hat{U}(t,t_0)\hat{b}_m\hat{
U}^{-1}(t,t_0)=\widehat{U(t,t_0)b_m}=
\alpha_m(t,t_0)\hat{b}_m+\beta_m(t,t_0)\hat{b}_{-m}^{\dag} \quad
\quad \forall m\in \mathbb{N}.
\end{equation}
In contrast with the situation found with the group of
$S^1$-translations, the complex structure $J_0$ is not invariant
under dynamical evolution, and hence neither is the cyclic vector
$|0\rangle$ \cite{cocome2,cocome}.

To conclude this section, let us remind that a symplectic
transformation $A$ is unitarily implementable on a Fock space
defined by a complex structure $J$ if and only if its antilinear
part $A_J=(A+JAJ)/2$ is Hilbert-Schmidt on the one-particle Hilbert
space defined by $J$ \cite{un1,un2}. An equivalent formulation is
that $J-AJA^{-1}$ be Hilbert-Schmidt. In the case of the considered
Fock representation for the Gowdy model, the condition of a unitary
implementation of the dynamics becomes
$\sum_{m=1}^{\infty}|\beta_m(t,t_0)|^2<\infty$, a finiteness that
was proved in \cite{cocome2,cocome}.

\section{Translation invariant complex structures}
\label{s1}

We now turn to the issue of determining the complex structures that
are invariant under the group of $S^1$-translations. We remember
that a complex structure $J$ is a {\em real} linear map on phase
space whose square is minus the identity. Therefore, $J$ must
commute with complex conjugation and $J^2=-{\bf 1}$. On the other
hand, $J$ must be compatible with the symplectic structure, namely,
$J$ must be a symplectic transformation and the bilinear map defined
on phase space by $\Omega(J\cdot,\cdot)$ must be positive definite,
so that $\{\Omega(J\cdot,\cdot)- i \Omega(\cdot,\cdot)\}/2$ provides
an inner product \cite{poincare}.

To these general properties we then add the following requirement.

\noindent {\bf Requirement 1} {\it
We consider only complex structures $J$ that are invariant under the
group of translations (\ref{4}), i.e. such that
$T_\alpha^{-1}JT_\alpha=J$ $\forall\alpha \in S^1$.}

This
requirement restricts considerably the admissible complex
structures, although the possible choices are still infinite.
We shall refer
to such complex structures simply as invariant ones.

\begin{prop} A compatible invariant complex structure $J$ is
necessarily block diagonal in the $\{B_m\}$ basis, each $4\times 4$
block being a matrix of the form
\begin{equation} \label{20} J_m= \left( \begin{array}{cc}
{\cal J}_m & {\bf 0}\\ {\bf 0} & {\cal J}_m
\end{array} \right),\quad\quad
{\cal J}_m= \left( \begin{array}{cc}
i\rho_m & \tilde{\rho}_m e^{i\delta_m} \\
\tilde{\rho}_m e^{-i\delta_m} & -i\rho_m
\end{array} \right),
\end{equation} where $\tilde{\rho}_m\geq 0$ and
$\rho_m=\sqrt{1+\tilde{\rho}_m^{\,2}}\geq 1$ $\forall m\in
\mathbb{N}$. The complex structure $J_0$ corresponds to
$\tilde{\rho}_m=0$ $\forall m\in \mathbb{N}$.
\end{prop}
{\bf Proof:} Employing transformations (\ref{8}) and (\ref{8b}), it
is straightforward to see that invariance under $S^1$-translations
requires a block diagonal form like that given in the first equation
in (\ref{20}), except for the fact that the two nonvanishing entries
may in principle be different $2\times 2$ matrices. Commutation with
complex conjugation allows then to express any of these matrices in
terms of the other. In addition, since $J$ is a complex structure,
$-J_m^2$ must equal the ($4\times 4$) identity matrix. Moreover,
compatibility with the symplectic structure (\ref{symstru}) implies
that $J_m^T\Omega_mJ_m=\Omega_m$ and
that $J_m^T\Omega_m$ must be positive definite. It is a simple
exercise to check that these conditions lead precisely to the above
general form for $J_m$.$\Box$

It turns out that the freedom in the choice of compatible invariant
complex structure is equivalent to that in performing a certain type
of symplectic transformations. More specifically, a direct
computation shows the following result.

\begin{prop} Every compatible invariant complex structure $J$ is
related to $J_0$ by a symplectic transformation $K_J$ (i.e.
$J=K_JJ_0K_J^{-1}$) that is block diagonal, with $4\times 4$ blocks
of the form
\begin{equation} \label{21} (K_J)_m= \left( \begin{array}{cc}
({\cal K}_J)_m & {\bf 0}  \\
{\bf 0} & ({\cal K}_J)_m
\end{array} \right),\quad\quad ({\cal K}_J)_m= \left(
\begin{array}{cc}\kappa_m & \lambda_m  \\ \lambda_m^* & \kappa_m^*
\end{array} \right),
\end{equation} where $|\kappa_m|^2-|\lambda_m|^2=1$.
Furthermore, there is a one-to-one correspondence between compatible
invariant complex structures and symplectic transformations of this
form with positive coefficients $\kappa_m$. Explicitly,
$\kappa_m=\sqrt{(\rho_m+1)/2}$ and
$\lambda_m=i\tilde{\rho}_me^{i\delta_m}/\sqrt{2(\rho_m+1)}$.
\end{prop}

Remember that the Fock representations defined by $J$ and $J_0$ are
equivalent if and only if $J-J_0$ is a Hilbert-Schmidt operator
on ${\cal H}_0$. In our case, using the above expressions, this
immediately translates into the condition
$\sum_{m=1}^{\infty}|\lambda_m|^2(1+2|\lambda_m|^2)< \infty$. This
is in turn equivalent to the summability of the sequence
$\{|\lambda_m|^2\}$. On the one hand, we have
$\sum_{m=1}^{\infty}|\lambda_m|^2\leq
\sum_{m=1}^{\infty}|\lambda_m|^2(1+2|\lambda_m|^2)$, so the former
of these sums is finite if so is the latter. On the other hand, if
$\sum_{m=1}^{\infty}|\lambda_m|^2<\infty$, all but at most a finite
number of elements in $\{|\lambda_m|^2\}$ are smaller than the
unity, so that $|\lambda_m|^4<|\lambda_m|^2$  for them, and hence
$\sum_{m=1}^{\infty}|\lambda_m|^2(1+2|\lambda_m|^2)$ must also be
finite.

In the following, we further restrict our attention to compatible
invariant complex structures $J$ that give rise to a unitary
implementation of the dynamics, i.e. such that the antilinear part
of the evolution operator, $\left\{U(t,t_0)+J U(t,t_0)J\right\}/2$,
is Hilbert-Schmidt with respect to the inner product
$\langle\cdot,\cdot\rangle_J=\left\{\Omega(J\cdot,\cdot)- i
\Omega(\cdot,\cdot)\right\}/2$ for all (strictly positive) values of
$t$ and $t_0$. It is straightforward to see that this
Hilbert-Schmidt condition can be reformulated as follows.

\begin{prop} Let $U$ be a symplectic transformation and $J$ and
$J_0$ two complex structures that are related by another symplectic
transformation $K_J$, $J=K_JJ_0K_J^{-1}$. Then the antilinear part
$\left(U+J U J\right)/2$ is Hilbert-Schmidt with respect to the
inner product $\langle\cdot,\cdot\rangle_J$ if and only if the $J_0$
antilinear part of $K_J^{-1}UK_J$, namely $\left(
K_J^{-1}UK_J+J_0K_J^{-1}U K_JJ_0\right)/2$, is Hilbert-Schmidt with
respect to $\langle\cdot,\cdot\rangle_{J_0}$.
\end{prop}

Applying this result to the symplectic transformation $U(t,t_0)$
provided by the evolution, the existence of a unitary implementation
of the dynamics with respect to $J$ becomes equivalent to that of a
unitary implementation of $U^J(t,t_0)=K_J^{-1}U(t,t_0)K_J$ with
respect to $J_0$ for all possible values of $t$ and $t_0$. Taken
then into account the general form of $U(t,t_0)$ and $K_J$, given in
equations (\ref{14}) and (\ref{21}), one gets an expression for
$U^J(t,t_0)$ which is again of the type (\ref{14}) but with
different coefficients $\alpha_m(t,t_0)$ and $\beta_m(t,t_0)$,
namely
\begin{equation} \label{22} U^J_m(t,t_0)= \left(
\begin{array}{cc}
{\cal U}^J_m(t,t_0) & {\bf 0} \\
{\bf 0} & {\cal U}^J_m(t,t_0)
\end{array} \right),\quad {\cal U}^J_m(t,t_0)= \left(
\begin{array}{cc}
\alpha^J_m(t,t_0) & \beta^J_m(t,t_0) \\
{\beta^J_m}^*(t,t_0) & {\alpha^J_m}^*(t,t_0)
\end{array} \right),
\end{equation}
with
\begin{eqnarray} \label{23}
\alpha^J_m(t,t_0)&=&|\kappa_m|^2\alpha_m(t,t_0)-|\lambda_m|^2
\alpha_m^*(t,t_0)+\kappa_m^*\lambda_m^*\beta_m(t,t_0)-\kappa_m
\lambda_m
\beta_m^*(t,t_0), \\
\label{24} \beta^J_m(t,t_0)&=&2i{\rm
Im}[\alpha_m(t,t_0)]\kappa_m^*\lambda_m+(\kappa_m^*)^2\beta_m(t,t_0)
-\lambda_m^2\beta_m^*(t,t_0).
\end{eqnarray}
Here, ${\rm Im}[z]$ denotes the imaginary part of $z$. Of
course, $U^J(t,t_0)$ is a symplectic transformation:
\begin{equation}\label{albeuno}|
\alpha^J_m(t,t_0)|^2-|\beta^J_m(t,t_0)|^2=
|\alpha_m(t,t_0)|^2-|\beta_m(t,t_0)|^2=1.\end{equation} The
condition for a unitary implementation of the dynamics in the $J$
representation is thus equivalent to the square summability of
$\{\beta^J_m(t,t_0)\}$, i.e. that
$\sum_{m=1}^{\infty}|\beta^J_m(t,t_0)|^2$ exists for all (strictly
positive) $t$ and $t_0$.

\section{Uniqueness of the complex structure}
\label{stuff2}

In this section we shall prove that any compatible invariant complex
structure $J$ that allows a unitary implementation of the dynamical
evolution provides a Fock representation that is unitarily
equivalent to that defined by $J_0$.

As we mentioned in the introduction, this proof applies not only to
the Gowdy model, but to a broader class of field dynamics. More
precisely, we consider classical evolutions of the form (\ref{9})
and (\ref{14}) such that the functions $\alpha_m(t_0+\tau,t_0)$
satisfy the following condition.

\noindent {\bf Condition 1} {\it There exist a (strictly positive)
value of $t_0$ and a constant $\delta\in(0,\pi)$ such that, for
every measurable set $\widetilde{E}_\delta\subset [0,\pi]$ with
Lebesgue measure $\mu(\widetilde{E}_\delta)>\pi-\delta$,
\begin{equation}\label{req2}\int_{\widetilde{E}_\delta}\left\{
1-({\rm Re}[\alpha_m(t_0+\tau,t_0)])^2
\right\}d\tau>\Delta(\widetilde{E}_\delta) \quad \quad \forall m\in
\mathbb{N}\end{equation} for certain strictly positive bounds
$\Delta(\widetilde{E}_\delta)>0$.}

Besides, we assume that the functions $\alpha_m(t_0+\tau,t_0)$ and
$\beta_m(t_0+\tau,t_0)$ are measurable functions of the elapsed time
$\tau=t-t_0$ on the closed interval $[0,\pi]$ for (at least) the
fixed value of $t_0$ given by condition 1. This additional
measurability condition is extremely mild and is obviously fulfilled
(for any choice of $t_0$) in the linearly polarised Gowdy
cosmologies [see definitions (\ref{15}) and (\ref{16})], as well as
in the case of the free massless field. On the other hand, we
postpone to section 5 the verification that condition 1 is satisfied
in the Gowdy model (and by the free massless field). Of course, we
are also assuming that the dynamics admits a unitary implementation
in the introduced representation $J_0$.

As stated above, we consider exclusively representations $J$ where
the evolution can also be implemented as a unitary transformation.
For an invariant representation of this kind, let us write
expression (\ref{24}) for $\beta^J_m(t,t_0)$ in a more convenient
form. We shall call $\phi_m$ and $\varphi_m$ the phases of
$\kappa_m$ and $\lambda_m$, respectively, and ${\tilde
\beta}_m(t,t_0)=\beta_m(t,t_0) e^{-i(\phi_m+\varphi_m)}$.
Multiplying equation (\ref{24}) by $e^{i(\phi_m-\varphi_m)}$, we
obtain after a trivial calculation
\begin{equation} \label{29} 2i|\kappa_m||\lambda_m|{\rm Im}[\alpha_m
+{\tilde \beta}_m]=e^{i(\phi_m-\varphi_m)}\beta^J_m-{\rm Re}[{\tilde
\beta}_m]- i(|\kappa_m|-|\lambda_m|)^2{\rm Im}[{\tilde \beta}_m],
\end{equation}
where ${\rm Re}[z]$ is the real part of $z$ and we have obviated the
time dependence. To further manipulate this relation, we shall use
the following (general) inequalities
\begin{eqnarray}\label{a)}
&&|y+z|^2\leq 2|y|^2+2|z|^2,\\
\label{b)} &&({\rm Im}[y+z])^2\geq \frac{a}{1+a}\left({\rm
Im}[z]\right)^2-a|y|^2 \quad \forall a\geq 0,\\
\label{c)} &&|\kappa_m|-|\lambda_m|\leq 1 \quad \forall m \in
\mathbb{N}.
\end{eqnarray}
The first one can be deduced by employing the triangle inequality
$|y+z|\leq |y|+|z|$. The second one is a consequence of the fact
that $(\sqrt{1+a}\,{\rm Im}[y]+{\rm Im}[z]/\sqrt{1+a}\,)^2\geq 0$.
Finally, the third one follows from the relation
$|\kappa_m|^2=1+|\lambda_m|^2$. Using these inequalities, together
with $|\kappa_m|\geq 1$ and
$|\alpha_m(t,t_0)|^2-|\beta_m(t,t_0)|^2=1$, it is not difficult to
show from equation (\ref{29}) that
\begin{equation} \label{30}
\frac{2a}{1+a}|\lambda_m|^2\left\{1-({\rm
Re}[\alpha_m(t,t_0)])^2-a|\beta_m(t,t_0)|^2\right\}\leq
2|\beta_m(t,t_0)|^2+|\beta^J_m(t,t_0)|^2
\end{equation}
for all (strictly positive) values of $t$ and $t_0$ and $\forall
m\in \mathbb{N},$ $a\geq 0$.

Let us substitute $t=t_0+\tau$ from now on and restrict our analysis
to the interval $\tau\in [0,\pi]$. In addition, let us sum equation
(\ref{30}) over $m$, from $m=1$ up to a finite but {\em generic}
$N\in\mathbb{N}$,
\begin{eqnarray} \label{38} \frac{2a}{1+a} &&\sum_{m=1}^N
|\lambda_m|^2\left\{1-({\rm
Re}[\alpha_m(t_0+\tau,t_0)])^2-a|\beta_m(t_0+\tau,t_0)|^2
\right\}\leq \nonumber\\ && \sum_{m=1}^N\left(2
|\beta_m(t_0+\tau,t_0)|^2+|\beta^J_m(t_0+\tau,t_0)|^2\right).
\end{eqnarray}

We now analyse the right hand side of this inequality. We restrict
all considerations to the fixed value of $t_0$ supplied by condition
1 and regard
\begin{equation} \sum_{m=1}^N|\beta_m(t_0+\tau,t_0)|^2 \quad {\rm
and} \quad \sum_{m=1}^N|\beta^J_m(t_0+\tau,t_0)|^2
\label{finsum}\end{equation} as functions of $\tau$ defined on
$[0,\pi]$. The imposed measurability condition on
$\alpha_m(t_0+\tau,t_0)$ and $\beta_m(t_0+\tau,t_0)$ together with
equations (\ref{23}) and (\ref{24}) guarantee that
$\alpha^J_m(t_0+\tau,t_0)$ and $\beta^J_m(t_0+\tau,t_0)$ are again
measurable functions of $\tau$ on $[0,\pi]$, and therefore the same
is true for the sums (\ref{finsum}) for every $N\in\mathbb{N}$. Note
also that the limit of these sums when $N\rightarrow \infty$ exists
for all values of $\tau$, because the dynamics is unitarily
implementable both in the $J_0$ and the $J$ representations (see end
of section 3). Moreover, we have the following integrability result.

\begin{lem}
\label{Luzin} For the considered fixed value of $t_0$ and $\forall
\delta>0$, there exist a measurable set $E_\delta\subset [0,\pi]$
with Lebesgue measure $\mu(E_\delta)>\pi-\delta$ and a positive
number $I_\delta$ such that
\begin{equation}
\int_{E_\delta}\sum_{m=1}^N\left(2|\beta_m(t_0+\tau,t_0)|^2
+|\beta^J_m(t_0+\tau,t_0)|^2\right)d\tau\leq I_\delta\quad \quad
\forall N\in\mathbb{N}.  \label{bound}\end{equation}
\end{lem}
{\bf Proof:} We present a proof that is based on Egorov's theorem,
which can be stated as follows (see e.g. \cite{kolfo}).
\begin{theo}{\bf (Egorov's theorem)} Let $E$ be  a set of finite
measure and $f_n(\tau)$ a sequence of measurable functions
converging (possibly only almost everywhere) to a function $f(\tau)$
on $E$. Then, $\forall \delta>0$, there exists a measurable set
$E_\delta\subset E$ with $\mu(E_\delta)>\mu(E)-\delta$ and such that
the convergence is uniform on $E_\delta$. \end{theo}

In our case $E=[0,\pi]$,
$f_N(\tau)=\sum_{m=1}^N\left(2|\beta_m(t_0+\tau,t_0)|^2+
|\beta^J_m(t_0+\tau,t_0)|^2\right)$ with fixed $t_0$ and
\begin{equation}f(\tau)=
2\sum_{m=1}^{\infty}|\beta_m(t_0+\tau,t_0)|^2+
\sum_{m=1}^{\infty}|\beta_m^J(t_0+\tau,t_0)|^2\label{sumf}.
\end{equation}
For every choice of $\delta>0$, let $E_\delta$ be as
in the theorem. We have
\begin{eqnarray} \nonumber \left|\int_{E_\delta}[f_{N_1}(\tau)
-f_{N_2}(\tau)]d\tau\right|&\leq
&\int_{E_\delta}|f_{N_1}(\tau)-f_{N_2}(\tau)|d\tau\leq
\pi\sup_{E_\delta}|f_{N_1}(\tau)-f_{N_2}(\tau)|\\
 &\leq & \pi \sup_{E_\delta}|f_{N_1}(\tau)-f(\tau)|
 +\pi\sup_{E_\delta}|f_{N_2}(\tau)-f(\tau)|,
\end{eqnarray} where we have employed that $\mu(E_\delta)<\pi$.
For any number $\epsilon>0$, the above upper bound is clearly
smaller than $\epsilon$ if $N_1$ and  $N_2$ are greater than a
certain integer $N_{\epsilon}$, because the convergence is uniform
on $E_\delta$. Thus, the sequence of integrals
$I_{\delta}^{(N)}=\int_{E_\delta}f_N(\tau)d\tau$ is Cauchy, and
therefore converges, i.e.
$I_\delta=\lim_{N\to\infty}I_{\delta}^{(N)}$ exists. Since the
partial sums $f_N(\tau)$ form an increasing sequence,
$f_{N+1}(\tau)\geq f_N(\tau)$, we have $I_{\delta}^{(N)}\leq
I_{\delta}^{(N+1)}$, and hence $I_{\delta}^{(N)}\leq I_\delta$
$\forall N\in\mathbb{N}$.$\Box$

We finally are in conditions to prove the uniqueness of the Fock
quantisation.

\begin{prop} \label{equiv} If condition 1 is
satisfied, $\sum_{m=1}^{\infty}|\lambda_m|^2$ is finite. \end{prop}
{\bf Proof:} We take values for $t_0$ and $\delta$ such that
condition 1 holds, and integrate both sides of relation (\ref{38})
over the corresponding set $E_\delta$ provided by lemma 1. Then,
condition 1 guarantees that inequality (\ref{req2}) is satisfied
with $\widetilde{E}_{\delta}=E_{\delta}$. Employing this and
equation (\ref{bound}) (which also provides the bound
$2\int_{E_\delta}|\beta_m(t_0+\tau,t_0)|^2d\tau\leq I_{\delta}$,
$\forall m\in \mathbb{N}$), we conclude that
\begin{equation}\label{proo}
\frac{a}{1+a}\sum_{m=1}^{N}|\lambda_m|^2\left[2\Delta(E_{\delta})
-aI_{\delta}\right]\leq I_{\delta}\quad \quad\forall a\geq
0,\quad\forall N\in\mathbb{N}.\end{equation} Choosing $a$ such that
$0<a<2\Delta(E_{\delta})/I_{\delta}$, we see that the partial sums
$\sum_{m=1}^N|\lambda_m|^2$ form a bounded increasing sequence:
\begin{equation}\label{proo2}
\sum_{m=1}^{N}|\lambda_m|^2\leq \frac{
(1+a)I_{\delta}}{a[2\Delta(E_{\delta}) -aI_{\delta}]}<\infty\quad
\quad\forall N\in\mathbb{N}.\end{equation} This guarantees that
$\sum_{m=1}^{\infty}|\lambda_m|^2$ is finite.$\Box$

Since the existence of this sum is the condition for unitary
equivalence of the Fock representations determined by the two
considered complex structures, our result proves that (modulo this
unitary equivalence) there is in fact a unique choice of compatible
invariant complex structure that permits the unitary implementation
of the dynamics when condition 1 (and the measurability condition on
the coefficients of the classical evolution operator) is satisfied.

\section{Completion of the proof for the Gowdy model}

To complete the proof of uniqueness of the Fock quantisation of the
linearly polarised Gowdy model, we still need to show that condition
1 is satisfied in this case. Remember that the functions
$\alpha_m(t_0+\tau,t_0)$ are given then by equation (\ref{15}). For
such functions, we have the following lemma.

\begin{lem}
\label{uni} Given any number $\epsilon\in(0,1]$, there exists a
constant $T_{\epsilon}>0$ such that, $\forall t_0>T_{\epsilon}$,
$\tau\in [0,\pi]$ and $m\in\mathbb{N}$,
\begin{equation} \label{31}
1-({\rm Re}[\alpha_m(t_0+\tau,t_0)])^2\geq \sin^2{(m\tau)}-
26\epsilon.
\end{equation}
\end{lem}
{\bf Proof:} Using expression (\ref{15}), it is straightforward to
see that
\begin{equation} \label{32} {\rm Re}[\alpha_m(t,t_0)]\leq
{\rm Re}[c(mt)c^*(mt_0)]+|d(mt)||d(mt_0)|
\end{equation}
with $t=t_0+\tau$. On the other hand, remembering definitions
(\ref{12}) and (\ref{13}) and the asymptotic behaviour of the Hankel
functions \cite{abra}, one can check that the functions $d(x)$ and
$C(x)=c(x)-e^{i\pi/4}e^{-ix}$ tend to zero when $x\rightarrow
\infty$. So, given any $\epsilon\in(0,1]$, there exists a constant
$T_{\epsilon}$ such that
\begin{equation}\label{boundepsil}\left|d(x)\right|\leq\epsilon,
\quad\quad |C(x)|=\left|c(x)-e^{i\frac{\pi}{4}}e^{-ix}\right|
\leq\epsilon\quad\quad \forall x>T_{\epsilon}.\end{equation} Since
$m\geq 1$ and $t=t_0+\tau\geq t_0$ with $\tau\in[0,\pi]$, the above
inequalities are valid $\forall t_0>T_{\epsilon}$ when $x$ equals
either $mt$ or $mt_0$.

In particular, $\forall t_0>T_{\epsilon}$, we obtain
\begin{eqnarray}\label{realp}
{\rm Re}[\alpha_m(t,t_0)]&\leq& {\rm Re}[c(mt)c^*(mt_0)]+\epsilon^2,\\
\label{ccine} \left|c(mt)c^*(mt_0)-e^{-im(t-t_0)}\right|^2&=&
\left|C(mt)
c^*(mt_0)+e^{-i\left(mt-\frac{\pi}{4}\right)}C^*(mt_0)\right|^2
\nonumber\\
&\leq& 2(|c^*(mt_0)|^2+1)\epsilon^2\leq 10 \epsilon^2.
\end{eqnarray}
In the last line we have used equation (\ref{a)}) and
$|c^*(mt_0)|^2\leq 4$, a bound that follows from equation
(\ref{boundepsil}) for $\epsilon\leq 1$. Direct consequences of
these inequalities are
\begin{eqnarray}\label{ineprim} {\rm Re}[c(mt)c^*(mt_0)]&\leq&
\cos{(m\tau)} +\sqrt{10}\epsilon\\ \label{reinequ}\left({\rm
Re}[\alpha_m(t,t_0)]\right)^2&\leq&
\cos^2{(m\tau)}+26\epsilon.\end{eqnarray} To arrive at this last
equation, we have employed relations (\ref{realp}) and
(\ref{ineprim}) and the fact that $|\cos{(m\tau)}|\leq 1$ and
$\epsilon^2\leq \epsilon$ when $0<\epsilon\leq 1$. The lemma follows
trivially from (\ref{reinequ}).$\Box$

Let us then take any number $\epsilon\in(0,1]$ and a fixed
$t_0>T_{\epsilon}$ according to lemma 2 and, for $\delta\in
(0,\pi)$, integrate relation (\ref{31}) over $\tau$ on arbitrary
sets $\widetilde{E}_{\delta}\subset [0,\pi]$ with
$\mu(\widetilde{E}_{\delta})>\pi-\delta$. Calling
$\overline{E}_\delta$ the complement of $\widetilde{E}_\delta$ with
respect to $[0,\pi]$, so that $\mu(\overline{E}_\delta)<\delta$, we
obtain $\forall m\in\mathbb{N}$:
\begin{eqnarray} \label{40} \nonumber \int_{\widetilde{E}_\delta}
\left\{1-({\rm Re}[\alpha_m(t_0+\tau,t_0)])^2 \right\}d\tau  &\geq &
\int_{\widetilde{E}_\delta} \sin^2{(m\tau)} d\tau- 26\epsilon\,\pi\\
&= &\frac{\pi}{2}- \int_{\overline{E}_\delta} \sin^2{(m\tau)} d\tau
-26\epsilon\,\pi \nonumber
\\&\geq &\frac{\pi}{2}-\delta
-26\epsilon\,\pi ,\end{eqnarray} where in the last step we have used
$\sin^2(m\tau)\leq 1$. We see that, as far as we choose
$\delta<\pi/2$ and $\epsilon<(\pi- 2\delta)/(52\pi)$, the sequence
of integrals (\ref{40}) $\forall m\in\mathbb{N}$ admit a strictly
positive lower bound for every choice of the set
$\widetilde{E}_{\delta}$ and for any choice of fixed value of
$t_0>T_{\epsilon}$. Therefore, condition 1 is indeed fulfilled
in the linearly polarised Gowdy model.

It is not difficult to realise that the basis of the above proof
resides in the fact that, owing to the asymptotic behaviour of the
Gowdy model at large times \cite{cocome2,cocome}, the coefficients
$\alpha_m(t_0+\tau,t_0)$ of the classical evolution operator in the
$\{B_m\}$ basis approach their counterparts for the free massless
scalar field in the limit $t_0\rightarrow\infty$. In the massless
case, the potential term $\xi/(4t^2)$ is absent in equation
(\ref{3}) and, as already mentioned,
\begin{equation} \alpha_m(t_0+\tau,t_0)=e^{-im\tau}
\quad \quad \forall m\in\mathbb{N}\label{free}.\end{equation} One
can see that these functions satisfy condition 1. Essentially, this
explains that the same occurs for the Gowdy model, taking into
account its asymptotic behaviour.

In more detail, direct substitution of expressions (\ref{free}) in
the integrals (\ref{40}) gives
\begin{eqnarray} \label{freeint} \nonumber
\int_{\widetilde{E}_\delta}\left\{1-({\rm
Re}[\alpha_m(t_0+\tau,t_0)])^2
\right\}d\tau &=&\int_{\widetilde{E}_\delta} \sin^2{(m\tau)} d\tau\\
&\geq &\frac{\pi}{2}- \delta\quad \quad\forall m\in\mathbb{N},
\end{eqnarray}
for all choices of $t_0$, $\delta\in(0,\pi)$ and
$\widetilde{E}_\delta$ with $\mu(\widetilde{E}_\delta)>\pi-\delta$.
Restricting $\delta$ to be smaller than $\pi/2$, we deduce the
existence of a strictly positive infimum for all wavenumbers $m$, so
that the condition is satisfied. Hence, our proof of uniqueness is
also valid for a free massless scalar field propagating in a flat
background with $S^1$ spatial sections.

\section{Conclusions and further comments}

In this work, we have analysed uniqueness criteria for the Fock
representation of a real scalar field satisfying a Klein-Gordon-like
equation in a flat 1+1 dimensional background with the spatial
topology of $S^1$, assuming that the field equations are invariant
under $S^1$-translations. We have proved that, if the complex
structure $^{\footnotemark[9]}$\footnotetext[9]{The complex and the
symplectic structures must be compatible, with the canonical
momentum of the field given by its time derivative.} is invariant
under the group of $S^1$-translations and allows a unitary
implementation of the dynamics, then the Fock representation is
unique up to unitary transformations, provided that the dynamics
satisfies certain conditions. In particular, these conditions
guarantee that the time average of some sequence of functions that
are related with the coefficients of the evolution operator
possesses a strictly positive infimum. We have also shown that these
conditions are fulfilled in the field description of the linearly
polarised Gowdy $T^3$ model introduced in \cite{cocome2,cocome}, as
well as in the case of the free massless scalar field.

The description of the Gowdy model formulated in
\cite{cocome2,cocome} involves an almost complete choice of gauge
(including deparameterisation) and a choice of field
parameterisation for the spatial metric. Our analysis demonstrates
that, once those choices have been made, the invariance under
$S^1$-translations and the unitary implementation of the evolution
pick out a unique Fock quantisation. Moreover, in the considered
description of the Gowdy model, the demand of invariance under
$S^1$-translations is well justified because these translations are
in fact a gauge group. Its generator corresponds then to a
constraint to be imposed {\it \`a la Dirac} on the kinematical Fock
space in order to arrive at the Hilbert space of physical states.
From this perspective, unitary implementation of the dynamics is
synonymous of uniqueness in the Fock quantisation of the linearly
polarised Gowdy $T^3$ cosmologies.

Our line of reasoning to prove the uniqueness of the Fock
representation has been the following. We have first shown that any
complex structure $J$ that is compatible with the symplectic form
and commutes with the group of $S^1$-translations has a very
specific block diagonal form in the $\{B_m\}$ basis (\ref{bcal})
(which is formed by the natural choice of annihilationlike and
creationlike variables for the case of the free massless scalar
field). We have then proved that all such complex structures can be
obtained by means of certain symplectic transformations from a
complex structure of reference, $J_0$ (namely, the structure which
would be selected by the energy condition of \cite{ash-mag} --or by
invariance under $S^1$-translations and dynamical evolution-- if the
scalar field were a free massless one). These symplectic
transformations have precisely the same type of block diagonal form
presented by the invariant complex structures. We have established a
one-to-one correspondence between a subset of such symplectic
transformations $K_J$ and the invariant complex structures $J$, so
that the choice of $K_J$ captures all the freedom available in the
construction of the Fock representation.

Using this result, we have reformulated the condition of unitary
implementation of the classical evolution operator $U$ in the
invariant representation $J$ as the unitary implementation of
$K_J^{-1}UK_J$ in the $J_0$ representation. Assuming this unitarity
and taking for granted that of $U$ in the $J_0$ representation (as
it is certainly the case for the Gowdy model and the free massless
field), we have arrived at an inequality that relates the antilinear
parts of $U$ and $K_J^{-1}UK_J$ with the antilinear part of $K_J$
[see equation (\ref{38})]. In this inequality, nonetheless, (the
square modulus of) the coefficients of the antilinear part of $K_J$
appear modulated by certain functions, determined by the classical
evolution operator $U$. These functions may in principle oscillate
and change their sign as the time $\tau$ elapsed from the Cauchy
surface of reference $t_0$ varies. To overcome this complication,
the idea is to average over $\tau$, eliminating in this way any
irrelevant oscillatory behaviour and local change of positivity of
the modulating functions.

Introducing the very mild assumption that the coefficients of the
classical evolution operator $U$ are measurable functions of $\tau$
on $[0,\pi]$ for fixed $t_0$, Egorov's theorem guarantees that there
exist subsets in that closed interval where the (trace of the square
norm of the) antilinear parts of $U$ and of $K_J^{-1}UK_J$ are
integrable. This assumption of measurability is satisfied both in
the free massless case and in the Gowdy model. Using this result,
uniqueness follows if, on any of those subsets of integrability and
for a suitable choice of the Cauchy surface $t_0$
$^{\footnotemark[10]}$,\footnotetext[10]{Different choices of
constant $t$-time Cauchy surfaces lead to unitarily equivalent
quantisations because the dynamical evolution between those surfaces
admits a unitary implementation in the $J_0$ representation.} the
corresponding time averages of all the modulating functions for the
different wavenumbers $m$ have a strictly positive infimum. This
last requirement on the dynamics ensures then that the antilinear
part of $K_J$ is Hilbert-Schmidt in the $J_0$ representation, so
that the symplectic transformation admits a unitary implementation.
As a consequence, the $J$ and $J_0$ representations turn out to be
unitarily equivalent.

Let us emphasise that, apart from the central role played by the
symmetry under $S^1$-translations, the only hypotheses made about
the details of the system are the unitary implementation of the
dynamics in the $J_0$ representation and the conditions that, at
some fixed value of $t_0$, the coefficients of the classical
evolution operator are measurable functions of $\tau$ on $[0,\pi]$
and the corresponding modulating functions present a strictly
positive infimum when averaged over $\tau$ (to be more precise, on
any subset of $[0,\pi]$ whose Lebesgue measure exceeds $\pi-\delta$
for certain constant $\delta>0$). We have verified these hypotheses
both in the linearly polarised Gowdy model and in the free massless
field. Actually, based on our proof that the averages have a
strictly positive infimum for the Gowdy model, one may convince
oneself that such a result can be generalised at least to those real
field dynamics where all the coefficients of the classical evolution
operator converge uniformly in $\tau$ and in the wavenumber $m$ to
their free massless counterparts, either at a fixed value of $t_0$
or asymptotically for infinitely large $t_0$. This includes, e.g.,
those dynamics that coincide with the free massless one in the
entire future of a Cauchy surface. In any of such circumstances, the
difference between the considered coefficients and their free
massless counterparts can be made as negligible as required for all
$\tau\in [0,\pi]$ and $m\in\mathbb{N}$ with a suitable choice of
$t_0$. The existence of a strictly positive lower bound (independent
of the wavenumber) on the averages for the system under study
follows then from that of the free massless field. Therefore,
uniqueness of the Fock quantisation holds also in these cases,
provided that the classical evolution operator admits a unitary
implementation in the considered ($J_0$) representation and that all
of its coefficients (in the $\{ B_m\}$ basis) are measurable
functions of $\tau\in [0,\pi]$ for the chosen value of $t_0$.

The compactness of the spatial sections of the background has
certainly been crucial in this proof of uniqueness. In particular,
the strict positivity of the infimum of the averaged modulating
functions is lost when the topology is noncompact. In that case, the
wavenumber $m$ would cease to be discrete, taking any positive
value. For instance, for the free massless field, whose modulating
functions are $\sin^2{m\tau}$, even the integral over the whole
interval $[0,\pi]$ becomes as small as desired when $m\rightarrow
0$, so that a strictly positive lower bound $\forall m>0$ does not
exist. The same would happen for the analogue of the Gowdy model
with noncompact spatial sections. In contrast, the dimension of the
spatial sections does not seem to play such a decisive role in our
discussion, in spite of the peculiarities that are usually tied with
symmetries in 1+1 dimensions. The possible generalisation of our
uniqueness result to higher dimensions will be the subject of future
research.

Another issue which deserves some comments is the choice of $J_0$ as
the complex structure of reference for the Fock quantisation of the
Gowdy model. As we have mentioned, this complex structure can be
regarded as the natural one associated with the free massless
dynamics. In principle, one might thought that an alternative
reasonable choice would be the complex structure $J_{M_{t_0}}$ that
corresponds (e.g. via the energy condition \cite{ash-mag}) to the
free dynamics with constant mass $M_{t_0}=1/(4t_0^2)$, namely, the
instantaneous value of the effective mass for the Gowdy model at the
chosen Cauchy surface $t_0$
$^{\footnotemark[11]}$\footnotetext[11]{As an aside, let us point
out that the complex structures that $J_{M_{t_0}}$ induces by time
evolution differ however from $J_{M_{t}}$.}. Nevertheless, the
complex structure $J_{M_{t_0}}$ presents the disadvantage of its
dependence on the choice of $t_0$. On the other hand, it actually
would lead to a unitarily equivalent Fock representation. Indeed,
one can check that $J_{M_{t_0}}-J_0$ is Hilbert-Schmidt in the $J_0$
representation if and only if the sequence $\{\lambda_m(M_{t_0})\}$
with $m\in \mathbb{N}$ and
\begin{equation}
\lambda_m(M_{t_0})=\frac{(m^2+M_{t_0}^2)^{1/4}}{\sqrt{m}}-
\frac{\sqrt{m}}{ (m^2+M_{t_0}^2)^{1/4}} \end{equation} is square
summable. This summability follows from that of $\{1/m^4\}$ taking
into account that $|\lambda_m(M_{t_0})|\leq (1+M_{t_0}^2)^{1/4}
M_{t_0}^2/(4m^2)$ $\forall m \in \mathbb{N}$ and $M_{t_0}\neq 0$.
The compact topology turns out again to be essential for the
equivalence of the representations determined by $J_0$ and
$J_{M_{t_0}}$, associated with different constant masses. For
instance, it is well known that different masses correspond to
inequivalent representations in the case of free scalar fields in
Minkowski spacetime \cite{RS2} (see also \cite{MTV} for a detailed
account of the role played in this respect by the long range
behaviour.).

Let us finally clarify that, as one would expect, the choice of the
complex structure $J_0$ for the quantisation of the Gowdy model is
in fact equivalent (in the sense of the unitary equivalence of the
Fock representations) to the prescription of \cite{ash-mag2}, which
appeals to the use of asymptotic complex structures. Since the
dynamics of the Gowdy model approaches that of the free massless
scalar field asymptotically (e.g., one may check in the Gowdy model
that the time average of the field energy on any interval
$[t_0,\infty)$ with $t_0>0$ equals the energy of the free massless
scalar field), it is possible to establish a symplectomorphism
between the spaces of smooth solutions for the Gowdy and the free
massless fields, respectively. Employing this symplectomorphism to
``pull back'' the natural complex structure of the massless case,
one obtains the following one for the Gowdy model \cite{ash-mag2}:
\begin{equation}
\tilde{J}_0=\lim_{t\rightarrow \infty} U(t_0,t)U_0(t,t_0)J_0
U_0(t_0,t)U(t,t_0)=\lim_{t\rightarrow \infty} U(t_0,t)J_0
U(t,t_0),\end{equation} where $U_0$ denotes the classical evolution
operator for the free massless dynamics. In the last identity, we
have employed that $J_0$ is invariant under such an evolution
operator. Making use of equations (\ref{12})-(\ref{16}), it is a
simple exercise to check that $\tilde{J}_0-J_0$ is indeed
Hilbert-Schmidt, so that both complex structures lead to equivalent
Fock representations.

\section*{Acknowledgements}

The authors are greatly thankful to A. Ashtekar for enlightening
conversations and suggestions. This work was supported by the joint
Spanish-Portuguese Project HP03-140, the Spanish MEC Project No.
FIS2005-05736-C03-02, the CONACyT U47857-F grant and the Portuguese
FCT Projects POCTI/FP/FNU/50226/2003 and POCTI/FIS/57547/2004. J.
Cortez was funded by the Spanish MEC, No./Ref. SB2003-0168.

\appendix

\section{The Hamiltonian and the vacuum}
\label{app-gt3-new}

In this appendix we show that a different way to guarantee the
uniqueness of the Fock representation defined by a compatible
invariant complex structure for the linearly polarised Gowdy model
(up to unitary equivalence) consists in replacing the condition of
unitary implementation of the dynamics by the stronger requirement
that the Fock vacuum belong to the domain ${\cal{D}}(\hat{H})$ of
the Hamiltonian operator.

In a system where time evolution is dictated by a self-adjoint
Hamiltonian operator $\hat{H}$, perturbative $S$-matrix analyses
cannot be performed outside ${\cal{D}}(\hat{H})$. In particular, in
a Fock representation of a field system, perturbative scattering
processes will be well defined on the dense subspace formed by the
states with a finite number of ``particles'' only if the vacuum
state $|0\rangle$ belongs to ${\cal{D}}(\hat{H})$. We do not
necessarily need to know how to calculate explicitly the action of
the evolution operator $\hat{U}$ on the whole Hilbert space, instead
we may just consider its series expansion and its action on (finite)
``$n$-particle'' states $|n\rangle$.

However, if the vacuum fails to be in the domain of the Hamiltonian,
one certainly cannot use the series expansion to evolve $|n\rangle$
states. A natural strategy to elude the technical complications
posed by this problem consists in searching for a unitary, time
independent Bogoliubov transformation that provides an alternative
(yet unitarily equivalent) Fock representation whose new vacuum
belongs to the domain of the Hamiltonian. When there exist classical
symmetries in the system, the requirement that they are unitarily
implemented restricts the possible Fock representations and
therefore the set of allowed Bogoliubov transformations. It might
happen that, among the infinitely many inequivalent Fock
representations of the field system, the demands of symmetry
invariance and a well defined action of the Hamiltonian on the
vacuum select just a single family of unitarily equivalent
representations. We shall see that this occurs with the quantum
description of the linearly polarised Gowdy model constructed in
\cite{cocome2,cocome}.

In that quantum description, and employing the $\{ B_m\}$ basis,
time evolution in the non-zero mode part of the field sector is
governed by the Hamiltonian \cite{cocome}:
\begin{equation}
\label{hamil-op} \hat{H}=\sum_{m=1}^{\infty}
\left\{\omega_{m}(t)\left[\hat{b}_{m}^{\dagger}\hat{b}_{m}
+\hat{b}^{\dagger}_{-m}\hat{b}_{-m}\right] + \rho_{m}(t)\left[
\hat{b}_{m}\hat{b}_{-m}+\hat{b}_{m}^{\dagger}\hat{b}_{-m}^{\dagger}
\right]\right\},
\end{equation}
where $\omega_{m}(t)=m+\rho_{m}(t)$ and $\rho_{m}(t)=1/(8mt^{2})$.
Since the sequence $\{\rho_{m}(t)\}$ is square summable (SS) for all
$t\in R^{+}$, the vacuum state belongs indeed to
${\cal{D}}(\hat{H})$.

Remembering that any compatible invariant complex structure $J$ can
be obtained from the complex structure $J_0$ by means of a
symplectic transformation $K_J$ of the form (\ref{21}), we can prove
the uniqueness of the Fock quantisation by showing that, if the
Bogoliubov transformation provided by $K_J$ leads to a new vacuum in
the domain of the Hamiltonian, then $K_J$ admits a unitary
implementation in the $J_0$ representation, i.e. $\{\lambda_{m}\}$
is SS. This Bogoliubov transformation is given by
\begin{equation}
\label{bogo-transf} b_{m}=\kappa_{m}a_{m}+\lambda_{m}a_{-m}^{*}\quad
\quad\forall m\in \mathbb{N}
\end{equation}
and a similar expression for $b_{-m}$ with $\kappa_{m}=\kappa_{-m}$
and $\lambda_{m}=\lambda_{-m}$.

In the new representation, the Hamiltonian operator adopts the form
\begin{equation}
\label{hamil-after-transf}
\hat{H}_J=\sum_{m=1}^{\infty}\left\{\eta_{m}(t)
\left[\hat{a}_{m}^{\dagger}\hat{a}_{m}+\hat{a}^{\dagger}_{-m}
\hat{a}_{-m}\right]+\gamma_{m}(t)\hat{a}_{m}\hat{a}_{-m}
+\gamma^{*}_{m}(t)\hat{a}^{\dagger}_{m}\hat{a}^{\dagger}_{-m}
\right\},
\end{equation}
where the time dependent coefficients are
\begin{eqnarray}
\eta_{m}(t) & = &
\omega_{m}(t)\left(|\kappa_{m}|^2+|\lambda_{m}|^2\right)
+\rho_{m}(t)\left(\kappa_{m}\lambda_{m}+\kappa^{*}_{m}
\lambda^{*}_{m}\right),
\label{inoc-term}\\
\gamma_{m}(t)&=&2\omega_{m}(t)\kappa_{m}\lambda^{*}_{m}
+\rho_{m}(t)\left[(\kappa_{m})^2+(\lambda_{m}^*)^2\right].
\label{impo-term}
\end{eqnarray}

Let us then assume that $\{\lambda_{m}\}$ is {\em not} SS but that
the ``new'' vacuum state $|0_J\rangle$ belongs to ${\cal
D}(\hat{H}_J)$, so that the sequence $\{\gamma_{m}(t)\}$ is SS. We
shall show that this leads to a contradiction. Writing
$\kappa_{m}=|\kappa_{m}|e ^{i\phi_{m}}$,
$\lambda_{m}=|\lambda_{m}|e^{i\varphi_{m}}$ and using the relation
$|\kappa_{m}|^2=1+|\lambda_{m}|^2$ and the expressions of
$\omega_{m}(t)$ and $\rho_{m}(t)$, we obtain
\begin{equation}
\label{theta-sq} |\gamma_{m}(t)|^2 =
\left[\Upsilon_{m}(t)\right]^2+\frac{1}{64m^2t^4}
\sin^{2}(\phi_{m}+\varphi_{m}),
\end{equation}
where $\Upsilon_{m}(t)\in {\mathbb{R}}$ is
\begin{equation}
\label{upsi}
\Upsilon_{m}(t)=2m|\kappa_{m}||\lambda_{m}|+\frac{1}{8mt^2}
\cos(\phi_{m}+\varphi_{m})+\frac{1}{4mt^2}
\left[|\kappa_{m}||\lambda_{m}|+|\lambda_{m}|^2
\cos(\phi_{m}+\varphi_{m})\right].
\end{equation}
The last term in equation (\ref{theta-sq}) defines a summable
sequence for all values of $t\in {\mathbb{R}}^{+}$, since
$\sin^{2}(\phi_{m}+\varphi_{m})\leq 1$ and the Riemann function
$Z(x)$ converges at $x=2$. Thus, the square summability of
$\gamma_{m}(t)$ amounts to that of $\Upsilon_{m}(t)$. We concentrate
our attention on the latter from now on.

Let $T>0$ be any strictly positive number and
\begin{equation}M_{T}=\left\{m\in {\mathbb{N}}:  \,
|\lambda_{m}|>\frac{1}{8mT^2}\right\}.\end{equation} Since we are
assuming that $\{\lambda_{m}\}$ is not SS, the set $M_T$ must
contain an infinite number of elements. For every $m\in M_{T}$,
\begin{equation}
\label{ineq1} 0<2m|\kappa_{m}||\lambda_{m}|-\frac{1}{8mT^2} \leq
2m|\kappa_{m}||\lambda_{m}|+\frac{1}{8mT^2}\cos(\phi_{m}+
\varphi_{m}),
\end{equation}
because $m|\kappa_m|>1$ and $|\cos{(\phi_{m}+ \varphi_{m}})|\leq 1$.
In addition, remembering that $|\lambda_m|\neq 0$ for $m\in M_{T}$,
we have
\begin{equation}
\label{ineq2} |\kappa_{m}||\lambda_{m}|+|\lambda_{m}|^2
\cos(\phi_{m}+\varphi_{m})>0.
\end{equation}
It then follows from definition (\ref{upsi}) that
\begin{equation}
\Upsilon_{m}(T)>2m|\kappa_{m}||\lambda_{m}|-\frac{1}{8mT^2}>0\quad
\quad\forall m\in M_T.\end{equation}

Moreover, employing again $m|\kappa_m|>1$ and
$|\lambda_{m}|>1/(8mT^2)$ for $m\in M_T$, one concludes from the
above inequality that
$\Upsilon_{m}(T)>(2m|\kappa_{m}|-1)|\lambda_{m}|>|\lambda_{m}|$. But
then the sequence $\{\Upsilon_{m}(T)\}$ with $m\in M_{T}$ cannot be
SS, because $\{\lambda_{m}\}$ with $m\in M_{T}$ is not. This clearly
implies that neither $\{\Upsilon_{m}(T)\}$ nor $\{\gamma_{m}(T)\}$
are SS when $m$ is allowed to run over the whole set $\mathbb{N}$,
in spite of our original assumptions. Therefore a contradiction
arises, signaling that if the vacuum state $|0_J\rangle$ is in the
domain of the Hamiltonian, the Bogoliubov transformation
(\ref{bogo-transf}) necessarily admits a unitary implementation (in
the $J_0$ representation). The two considered representations are
thus unitarily equivalent, as we wanted to show. Let us finally
stress that unitarity of the Bogoliubov transformation
(\ref{bogo-transf}) is just a necessary condition, but not a
sufficient one in order to ensure that the state $|0_J\rangle$
belongs to ${\cal D}(\hat{H}_J)$.


\begin{thebibliography}{99}

\bibitem{wald} Wald R M 1994 {\em Quantum Field Theory in
Curved Spacetime and Black Hole Thermodynamics} (Chicago: Chicago
University Press)

\bibitem{SVN} Simon B 1972 {\em Topics in Functional Analysis} ed
R F Streater (London: Academic Press)

\bibitem{ash-lew-rep} Ashtekar A and Lewandowski J 2004
Background independent quantum gravity: A status report {\em Class.
Quantum Grav.}  {\bf 21} R53
%%CITATION = GR-QC 0404018;%%

\bibitem{rov} Rovelli C 2004 {\em Quantum
Gravity} (Cambridge, UK: Cambridge University Press)

\bibitem{thie}
Thiemann T 2001 Introduction to modern canonical quantum general
relativity {\em Preprint} gr-qc/0110034
%%CITATION = GR-QC 0110034;%%

\bibitem{lost} Lewandowski J, Okolow A, Sahlmann H and Thiemann T
2005 Uniqueness of diffeomorphism invariant states on holonomy-flux
algebras {\em Preprint} gr-qc/0504147
%%CITATION = GR-QC 0504147;%%

\bibitem{poincare} See e.g. Corichi A, Cortez J and Quevedo H
2004 Schr\"odinger
and Fock representation for a field theory on curved spacetime {\em
Ann. Phys.} {\bf 313} 446
%%CITATION = HEP-TH 0202070;%%

\bibitem{ash-mag} Ashtekar A and Magnon A 1975 Quantum
fields in curved space-times {\em Proc. R. Soc. Lond.} A
{\bf 346} 375\\
%%CITATION = PRSLA,A346,375;%%
Ashtekar A and Magnon-Ashtekar A 1980 A curiosity concerning the
role of coherent states in quantum field theory {\em Pramana} {\bf
15} 107
%%CITATION = PRAMC,15,107;%%

\bibitem{misner1} Misner C W 1972 Minisuperspace
{\em Magic without Magic: John Archibald Wheeler} ed J Klauder (San
Francisco: Freeman)

\bibitem{midi-sm} Torre C G 1999 Midisuperspace models of
canonical quantum gravity {\em Int. J. Theor. Phys.} {\bf 38} 1081
%%CITATION = GR-QC 9806122;%%

\bibitem{cocome2} Corichi A, Cortez J and Mena Marug\'an G A 2006
Unitary evolution in Gowdy cosmology {\em Phys. Rev.} D {\bf 73}
041502
%%CITATION = GR-QC 0510109;%%

\bibitem{cocome} Corichi A, Cortez J and Mena Marug\'an G A 2006
Quantum Gowdy $T^3$ model: A unitary description {\em Phys. Rev.} D
{\bf 73} 084020
%%CITATION = GR-QC 0603006;%%

\bibitem{gowdy} Gowdy R H 1974 Vacuum space-times with two
parameter spacelike isometry groups and compact invariant
hypersurfaces: topologies and boundary conditions {\em Ann. Phys.}
{\bf 83} 203
%%CITATION = APNYA,83,203;%%

\bibitem{misn} Misner C W 1973 A minisuperspace example:
The Gowdy $T^3$ cosmology {\em Phys. Rev.} D {\bf 8} 3271

\bibitem{berger} Berger B K 1974 Quantum graviton creation in a
model universe {\em Ann. Phys.} {\bf 83} 458
\\
Berger B K 1975 Quantum cosmology: Exact solution for the Gowdy
$T^3$ model {\em Phys. Rev.} D {\bf 11} 2770
\\
%%CITATION = PHRVA,D11,2770;%%
Berger B K 1984 Quantum effects in the Gowdy $T^3$ cosmology {\em
Ann. Phys.} {\bf 156} 155
%%CITATION = APNYA,156,155;%%

\bibitem{hs} Husain V and Smolin L 1989 Exactly
solvable quantum cosmologies from two Killing field reductions of
general relativity {\em Nucl. Phys.} B {\bf 327} 205
%%CITATION = NUPHA,B327,205;%%

\bibitem{guillermo}
Mena Marug\'an G A 1997 Canonical quantization of the Gowdy model
{\em Phys. Rev.} D {\bf{56}} 908
%%CITATION = GR-QC 9704041;%%

\bibitem{pierri} Pierri M 2002 Probing quantum general relativity
through exactly soluble midi-superspaces. II: Polarized Gowdy models
{\em Int. J. Mod. Phys.} D {\bf 11} 135
%%CITATION = GR-QC 0101013;%%

\bibitem{ccq-t3} Corichi A, Cortez J and Quevedo H 2002
On unitary time evolution in Gowdy $T^3$ cosmologies {\em Int. J.
Mod. Phys.} D {\bf 11} 1451
%%CITATION = GR-QC 0204053;%%

\bibitem{come} Cortez J and Mena Marug\'an G A 2005 Feasibility of a
unitary quantum dynamics in the Gowdy $T^3$ cosmological model {\em
Phys. Rev.} D {\bf 72} 064020
%%CITATION = GR-QC 0507139;%%

\bibitem{torre-prd} Torre C G 2002 Quantum dynamics of the polarized
Gowdy $T^3$ model {\em Phys. Rev.} D {\bf 66} 084017
%%CITATION = GR-QC 0206083;%%

\bibitem{jacob} Jacobson T 1991 Unitarity, causality
and quantum gravity {\em Conceptual Problems of Quantum Gravity} ed
A Ashtekar and J Stachel (Boston: Birkh\"auser)

\bibitem{macca} Kramer D, Stephani H, MacCallum M and Herlt E 1980
{\em Exact Solutions of Einstein's Field Equations} (Cambridge, UK:
Cambridge University Press)

\bibitem{choma} Cho D H J and Varadarajan M 2006 Functional
evolution of quantum cylindrical waves {\em Preprint} gr-qc/0605065

\bibitem{abra} Abramowitz M and Stegun I A (ed) 1970
{\em Handbook of Mathematical Functions} 9th edn. (Nat. Bur. Stand.
Appl. Math. Ser. No. 55) (Washington D.C.: U.S. Govt. Print Off.)

\bibitem{fock} See e.g.
Choquet-Bruhat Y, DeWitt-Morette C and Dillard-Bleick M 1996 {\em
Analysis, Manifolds and Physics, Part I: Basics} (Amsterdam:
Elsevier)
\\
Kay B S 1978 Linear spin-zero quantum fields in external
gravitational and scalar fields I. A one particle structure for the
stationary case {\em Commun. Math. Phys.} {\bf 62} 55
%%CITATION = CMPHA,62,55;%%

\bibitem{un1}  Shale D 1962 Linear symmetries of free boson
fields {\em Trans. Am. Math. Soc.} {\bf 103} 149

\bibitem{un2} Honegger R and Rieckers A 1996 Squeezing Bogoliubov
transformations on the infinite CCR-algebra {\em J. Math. Phys.}
{\bf 37} 4292

\bibitem{kolfo} Kolmogorov A N and Fomin S V 1999 {\em Elements of
the Theory of Functions and Functional Analysis} (New York: Dover)

\bibitem{RS2} Reed M and Simon B 1975 {\em Methods of Modern
Mathematical Physics II: Fourier Analysis, Self-Adjointeness} (San
Diego: Academic Press)

\bibitem{MTV} Mour\~ao J M, Thiemann T and Velhinho J M 1999
Physical properties of  quantum field theory measures {\em J. Math.
Phys.} {\bf 40} 2337
%%CITATION = HEP-TH 9711139;%%

\bibitem{ash-mag2} Ashtekar A and Magnon-Ashtekar A 1980 A
geometrical approach to external potential problems in quantum field
theory {\em Gen. Rel. Grav.} {\bf 12} 205
%%CITATION = GRGVA,12,205;%%

\end{thebibliography}
\end{document}